# Enhancement of spin relaxation time in hydrogenated graphene spin valve devices

M. Wojtaszek,[1] I. J. Vera-Marun,[1] T. Maassen,[1] and B. J. van Wees[1]

[1]*Physics of Nanodevices, Zernike Institute for Advanced Materials,
University of Groningen, Groningen, The Netherlands*[*]

(Dated: August the 28[th], 2012)

Hydrogen adsorbates in graphene are interesting as they are not only strong Coulomb scatterers but they also induce a change in orbital hybridization of the carbon network from sp$^2$ into sp$^3$. This change increases the spin-orbit coupling and is expected to largely modify spin relaxation. In this work we report the change in spin transport properties of graphene due to plasma hydrogenation. We observe an up to three-fold increase of spin relaxation time $\tau_S$ after moderate hydrogen exposure. This increase of $\tau_S$ is accompanied by the decrease of charge and spin diffusion coefficients, resulting in a minor change in spin relaxation length $\lambda_S$. At high carrier density we obtain $\lambda_S$ of 7 $\mu$m, which allows for spin detection over a distance of 11 $\mu$m. After hydrogenation a value of $\tau_S$ as high as 2.7 ns is measured at room temperature.

PACS numbers: 72.25.-b, 85.75.-d, 68.43.-h

Spin transport in graphene attracts a lot of research attention due to its potential for high spin relaxation times $\tau_S$ and large spin relaxation lengths $\lambda_S$ in supported[1–3] and suspended[4] samples at room temperature. These remarkable spin transport properties of graphene originate from a very low intrinsic spin-orbit coupling ($\Delta_{SO} \approx 1\mu eV$[5,6]), small hyperfine interactions (a low abundance of the C$^{13}$ isotope) and the low atomic number of carbon atoms. Still the experimentally observed spin relaxation length of a few micrometers is much below the theoretical limit[7,8]. The two main relevant mechanisms of spin relaxation[9,10] are: (1) Elliott-Yafet, where $\tau_S$ is directly proportional to momentum scattering time $\tau_P$: $\tau_S \sim \tau_P$, and (2) D'yakonov-Perel, where these times are inversely proportional $\tau_S \sim 1/\tau_P$. For both mechanisms $\tau_S$ decreases when the Rashba spin-orbit coupling increases. This can happen due to ripples[5], perpendicular electric fields[5,11] or adsorbed adatoms[6].

Hydrogen adsorbates in graphene are mostly studied in charge transport measurements[12,13] but they are also interesting for spin transport properties. These defects induce tetragonal distortion and sp$^3$ hybridization of the atomic orbitals in graphene, which strongly enhances the spin-orbit coupling ($\Delta_{SO} \approx 7$ meV)[6]. Similar to other point defects in the graphene lattice (vacancies, C, N adatoms), hydrogen adatoms create a localized state around the Dirac point in the electronic band structure and induce magnetic moments of $\sim 1\mu_B$ in the surrounding carbon atoms, where $\mu_B$ is the Bohr magneton[14,15]. These magnetic properties of adatoms lead to the observation of the tunable Kondo effect[16], spin-half paramagnetism[17,18] and recently modulation of spin transport[19] at low temperatures.

In this paper we investigate the change in the room temperature spin transport properties of graphene due to plasma hydrogenation. We characterize charge ($D_C$) and spin ($D_S$) diffusion coefficients, spin relaxation time $\tau_S$ and spin relaxation length $\lambda_S$ using Hanle precession measurements before and after hydrogenation in two separately fabricated devices. Upon hydrogenation, we observe a decrease of the diffusion coefficient and simultaneously the enhancement of $\tau_S$. This opposite trend results in only moderate change of $\lambda_S$ (within 30% of initial value). After hydrogenation we measure $\tau_S \approx 2.7$ ns at high carrier concentration n = 5.2x10$^{12}$ cm$^{-2}$, which is a record value at room temperature for single layer graphene. We discuss how different properties of hydrogen adatoms can lead to such an enhancement of $\tau_S$.

The graphene spin valve devices are fabricated by mechanical exfoliation of highly oriented pyrolytic graphite onto 500 nm SiO$_2$ with a highly-doped Si substrate below, acting as a gate electrode. Single layer graphene is selected based on the optical contrast, which was calibrated beforehand with Raman spectroscopy[20]. Contacts are fabricated by electron beam lithography, evaporated and shaped in a lift-off process. Materials are evaporated as follows: first 0.8 nm of Al with a deposition rate of 1 Å/s, which we oxidize in ambient air, so that it acts as a tunneling barrier. Then we evaporate Co (30 nm) and 2 nm of Al on top as a capping layer to protect it from the air oxidation and plasma treatment. The tunneling barrier of AlO$_x$ is present only underneath the contacts.

Chemisorption of hydrogen is realized *ex-situ* at room temperature in a reactive ion etching setup, where the whole device is exposed to Ar/H$_2$ plasma (85:15%) using exactly the same parameters as in Ref. 21. Hydrogen plasma exposure results in the development of a D band at 1340 cm$^{-1}$ in the graphene Raman spectrum, a signature of sp$^3$ defects, and in the increase of electrical resistivity, with a maximum well above 5 k$\Omega$. For accurately tuned plasma conditions (low plasma ignition power, high gas pressure in the process chamber, zero self-bias) little or no vacancies are created despite the fact that the topography of graphene becomes more rippled[21,22]. Each plasma exposure step lasts 10 minutes and is followed by electronic measurements at room temperature in vacuum. Although for devices studied here no Raman spectra is obtained, a roughly threefold increase in the sheet resistivity of graphene, as shown in Fig. 1, can be only attributed to hydrogenation.

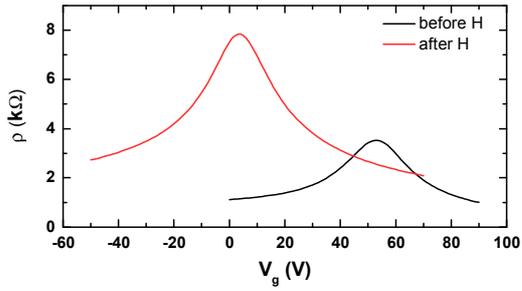

FIG. 1: (Color online) Sheet resistivity of graphene before and after plasma hydrogenation in device A.

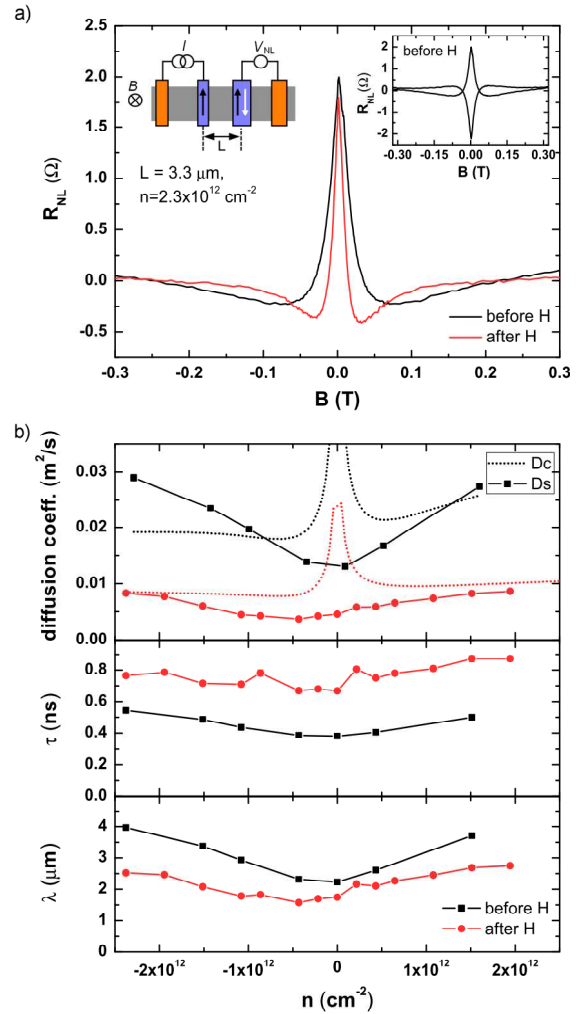

FIG. 2: (Color online) (a) Hanle precession measurements in device A before and after plasma hydrogenation. After hydrogenation the width of the Hanle curve decreases indicating the increase of spin relaxation time and/or decrease of diffusion coefficient, both observed here. The left inset shows a schematic of a non-local measurement, the right inset shows a measurement in parallel (P) and antiparallel (AP) configuration before hydrogenation. (b) Comparison of transport properties before and after hydrogenation as a function of carrier concentration. Top panel: diffusion coefficient, central panel: spin relaxation time $\tau_S$, bottom panel: spin relaxation length $\lambda_S$. $D_C$ is calculated using the Einstein relation (charge transport), $D_S$ is obtained from Hanle precession (spin transport).

We characterize spin transport before and after hydrogenation in two separately fabricated devices. In device A, in which the outermost contacts are non-magnetic, we perform one hydrogenation step; in device B, with all contacts magnetic, we perform two hydrogenation steps. As we measure different doping regimes in each of the devices, we discuss them separately, however both display a similar trend upon hydrogenation.

Before and after every hydrogenation step we characterize the sheet resistance $\rho$ of graphene as a function of the gate bias $V_g$ in a 4 terminal measurement. The induced carrier concentration $n$ is calculated from: $n = C_g/e(V_D - V_g)$, where $V_D$ is the voltage corresponding to the maximum of $\rho$, which is called charge neutrality point or Dirac point, and $C_g$ is the gate capacitance, $C_g = 70$ aF/$\mu m^2$ for 500 nm $SiO_2$.

Before hydrogenation in device A $V_D = 55$ V, indicating a p-doping of the graphene flake, which usually originates from lithographic processing of the sample, see Fig. 1. After hydrogenation $V_D$ reduces to 4 V, mainly due to the change of the work function of plasma hydrogenated graphene with respect to the $SiO_2$ substrate[23].

After hydrogenation the sheet resistivity of graphene increases roughly by a factor of three and as a consequence the mobility decreases from ∼3000 to 1000 cm$^2$/Vs. Considering the change in the mean free path of graphene before versus after hydrogenation and an electronic scattering cross section of 7 nm, as determined in Ref. 21, we estimate the hydrogen coverage to be 0.02% (4.5x10$^{11}$ atoms/cm$^{-2}$). Such a low level of hydrogenation does not change the graphene DOS significantly, the transport remains in the diffusive regime and one can reduce the role of hydrogen adatoms to charge/spin scattering centers.

Further, we perform spin transport measurements in a non-local spin valve geometry, as depicted schematically in the left inset of Fig. 2a. In there we separate the charge from the spin current using the fact that while injected charge current $I$ is only present between the injector and the drain, the injected spin current diffuses in all directions and can be detected as the spin voltage drop $V_{NL}$ in another part of the device. A non-local spin resistance, defined as $R_{NL} = V_{NL}/I$, displays a switching 'spin-valve' behavior in an in-plane magnetic field, depending on the relative magnetization alignment of the injecting and detecting magnetic contacts, see Supp. Inf.

To get a complete overview of spin transport properties before and after hydrogenation we perform Hanle precession measurements[1,24] for a set of gate voltages and a set of distances $L$ between the injector and detector. In these measurements the magnetic field is applied perpendicular to the graphene device, so that the injected spins having in-plane magnetization start to precess around the vector normal to the graphene plane. To account for the common background we measure the precession for parallel (P) and antiparallel (AP) configuration of the contact magnetization, like in the right inset of Fig. 2a.

The spin dynamics are described by the 1 dimensional Bloch equation for the spin chemical potential $\vec{\mu_S}$:

$$D_S \nabla^2 \vec{\mu_S} - \frac{\vec{\mu_S}}{\tau_S} + \vec{\omega_0} \times \vec{\mu_S} = \vec{0}. \tag{1}$$

which includes spin diffusion: the term with $D_S$, spin relaxation: the term with $\tau_S$ and spin precession due to external magnetic field $\vec{B}$: the term with the Larmor frequency $\vec{\omega_0} = g\mu_B \vec{B}/\hbar$, where $g = 2$ is the gyromagnetic factor of a free-electron. By fitting the solution of the Bloch equation to the precession measurements the spin coefficients $\tau_S$, $D_S$, $\lambda_S$, where $\lambda_S = \sqrt{D_S \tau_S}$, are obtained.

An example of Hanle measurements before and after hydrogenation is presented in Fig. 2a. Upon hydrogenation the maximum of $R_{NL}$ at zero magnetic field decreases only slightly, however the full width at half maximum (FWHM) of the precession curve reduces from 24 mT to 14 mT. This indicates an increase of $\tau_S$ or a decrease in $D_S$ and both of these effects are observed here: $\tau_S$ changes from 0.5 to 0.7 ns, whereas $D_S$ changes from 0.03 to 0.01 m$^2$/s.

The extracted spin coefficients at different doping are summarized in Fig. 2b. The diffusion coefficient is independently extracted from charge ($D_C$) and from spin ($D_S$) transport. $D_C$ is calculated using the Einstein relation $\sigma = e\nu(E)D_C$, where $\nu(E)$ is the density of states at T = 0 K and $\sigma = 1/\rho$ is its sheet conductivity. The singularity of $D_C$ around $n = 0$ comes from the vanishing number of states at the Dirac point and can be eliminated by including a broadening of $\nu(E)$ of graphene[25]. In our case, a good agreement between $D_C$ and $D_S$ is obtained when one uses the broadening factor $\tilde{\sigma} = 110$ meV, as analyzed in Supp. Inf. For pristine graphene $\tau_S$, $D_S$ and $\lambda_S$ are increasing with carrier concentration, like in previous reports[2,25]. After hydrogenation $\tau_S$ increases at all measured dopings, approaching $\tau_S = 1$ ns for $n = 2$x$10^{12}$ cm$^{-2}$, see Fig. 2b. Along with the decrease of electron mobility, $D_C$ and $D_S$ are strongly reduced. The opposite change in $D_S$ when compared with $\tau_S$ upon hydrogenation results in a slight decrease of $\lambda_S$ (here by 30%). The spin transport properties of the device are similar for a range of different distances between injector and detector $L$=1.3, 3.3 and 5.5 $\mu$m and the electrically floating contacts in between do not introduce any observable extra spin relaxation (see Supp. Inf.).

A very similar behavior of $\tau_S$ and $D_S$ upon hydrogenation is measured in device B. In there we measured the spin transport for injector-detector spacing $L$= 1.5, 3, 4, 6.5, 7, 8 and 11 $\mu$m. After two hydrogenation steps we obtain $\tau_S \approx 2.7$ ns at $n = 5.2$x$10^{12}$ cm$^{-2}$, which is the highest carrier concentration measured here. The Hanle data for this case, together with the fits are presented in Fig. 3a. The high value of $\tau_S$ contributes to the long spin relaxation length of $\lambda_S = 7$ $\mu$m, which is the largest value reported so far for graphene on SiO$_2$ at room temperature in a non-local measurement. A such high value of $\lambda_S$ is consistent with our observation of the spin signal over a distance of $L$=11 $\mu$m.

The behavior of the spin coefficients in device B for the initial state and two consecutive hydrogenation steps is presented for three different carrier concentrations in Fig. 3 (b, c, d) for $L = 6.5$ $\mu$m and like in the previous device the spin coefficients obtained from measurements over different $L$ are similar (see Supp. Inf.). After every hydrogenation step $\tau_S$ increases for all carrier concentrations and $D_S$ decreases, like in device A. The small upshift of $D_S$ after the second hydrogenation step is attributed to the uncertainty in determining $n$. Device B shows a drift of the Dirac point in time, hence the carrier concentration is determined with an accuracy of $\Delta n$ =5x10$^{11}$ cm$^{-2}$. $\lambda_S$ shows minor changes (less than 30%) after hydrogenation, matching the behaviour of device A. We note that the strength of hydrogenation depends on the individual properties of the graphene flake (its surface roughness, the amount of surface residues etc.[26,27]) so the comparison between device A and B can be only qualitative.

The proportionality between $\tau_S$ and $\tau_P$ in pristine graphene[25] was initially interpreted as a proof of the EY scattering mechanism being the dominant one in graphene. However, recent theoretical works[28,29] showed that for the EY mechanism in graphene this proportionality depends also on the Fermi energy $E_F$: $\tau_S \sim E_F^2 \tau_P$, which is not observed when varying the carrier concentration. Moreover, when one accounts for charged adatoms, substrate induced ripples and random Rashba fields[29] DP mechanism can exhibit EY-like or DP-like dependence. As it is not trivial to differentiate between the EY and DP mechanism it is usually assumed that both mechanisms are simultaneously present in graphene. The increase of $\tau_S$ with the decrease of $\tau_P$ upon hydrogenation suggests at first a DP spin relaxation mechanism. However, hydrogen sp$^3$ defects should enhance spin-orbit coupling and hence reduce $\tau_S$ irrespectively to the details of the spin scattering mechanisms (DP or EY[29,30]). One possible explanation could be that hydrogenation modifies the built-in spin-orbit coupling, reducing its intrinsic effectiveness and that property in combination with DP scattering mechanism results in an increase of $\tau_S$. A hydrogen defect, however, is said to not only enhance spin-orbit coupling but also to produce a localized state and a magnetic dipole moment in its vicinity[6,14] and all these three effects should be accounted for.

Recently, a large difference between the diffusion coefficient obtained from charge and from spin transport, where $D_S$ was $\geq$50 times lower than $D_C$, was observed in epitaxial graphene on SiC(0001)[31]. This discrepancy can possibly be explained by the influence of the localized states from the buffer layer between the graphene and the SiC substrate[32], which trap spins and cause the decrease of $D_S$ while increasing $\tau_S$. Such property of localized states might also be relevant for hydrogenated graphene.

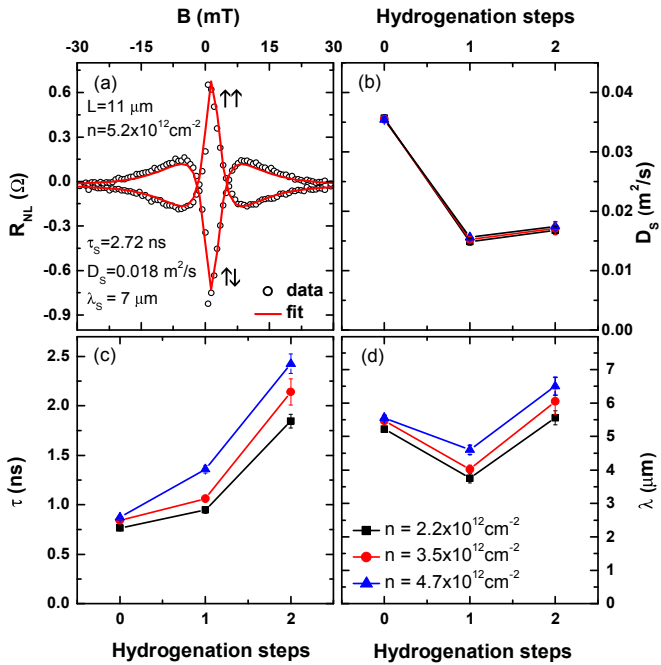

FIG. 3: (Color online) (a) Example of raw data of P and AP Hanle precession curves for device B after two hydrogenation steps and their fits to the solution of the Eq. 1. The extracted coefficients: $\tau_S \approx 2.7$ ns and $\lambda_S = 7$ μm have the largest values reported so far for single layer graphene on $SiO_2$ at room temperature. (b-d) Spin transport properties, $\tau_S$, $D_S$ and $\lambda_S$ in device B at different hydrogenation steps for three different carrier concentrations and $L=6.5$ μm. A large increase in $\tau_S$, decrease in $D_S$ and minor change in $\lambda_S$ with hydrogenation are observed, similar to device A.

On the other hand, the infuence of localized magnetic moments of hydrogenated and Ar-sputtered graphene was recently addressed in a low temperature study and a phenomenological theory for spin scattering was proposed[19]. In that paper the three key experimental features were explained by the presence of diluted magnetic moments: (1) a dip in $R_{NL}$ around zero in-plane magnetic field, originating from fluctuating magnetic moments, (2) narrowing of the Hanle precession curves and (3) a large discrepancy between $D_C$ and $D_S$. The last two were related to the change in the $g$-factor or equivalently the change in effective magnetic field due to the contribution from magnetic defects. This change led to discrepancies between real and fitted values of $\tau_S$ and $D_S$ in Hanle precession measurements. As the solution of the Bloch equation is invariant under the transformation $g \to cg^*$, $\tau_S \to \tau_S^*/c$, $D_S \to cD_S^*$ we are unable to uniquely determine all three variables. When $c > 1$ the extracted values of $\tau_S$ are overestimated and values of $D_S$ are underestimated comparing to the intrinsic values $\tau_S^*$ and $D_S^*$, but their product remains the same. As $\lambda_S = \sqrt{\tau_S D_S} = \sqrt{\tau_S^* D_S^*}$ independent on the actual transformation, the extracted value here $\lambda_S = 7$ μm after hydrogenation is reliable.

In our spin-valve measurements we do not observe a clear reduction of $R_{NL}$ around zero in-plane magnetic field[19,33], see Fig. 1 in Supp. Inf. Also, the Hanle precession curves after hydrogenation fit well to the solutions of Eq. 1 in the whole range of magnetic field. Finally, the extracted values of $D_S$ are very similar to $D_C$, which cannot be the case if the $g$-factor has substantially changed due to paramagnetic defects, see Supp. Inf.. From these considerations we conclude that the possible magnetization from hydrogen defects does not introduce any significant effective magnetic field into Hanle precession measurements and the obtained values of $\tau_S$ and $D_S$ are reliable.

The presented study of graphene spin transport with hydrogen defects forms a bridge between the low temperature study of graphene with local magnetic dipoles, when one ignores localized states[19] and the room temperature study of SiC graphene with non-magnetic localized states in its close proximity (buffer layer)[31,32].

This paper discusses changes in charge and spin transport for graphene after introducing hydrogen defects of low concentration at room temperature. The measured spin valve devices have initially good spin transport properties, with spin relaxation times above 0.5 ns. After plasma hydrogenation we observe a large increase of the spin relaxation time $\tau_S$ and a decrease of the diffusion coefficient extracted independently from charge ($D_C$) and spin ($D_S$) transport measurements, with a minor effect on the spin relaxation length $\lambda_S$. After two hydrogenation steps we measure $\tau_S = 2.7$ ns and $\lambda_S = 7$ μm at $n = 5.2 \times 10^{12}$ cm$^{-2}$, which are the highest room temperature values reported so far for graphene on $SiO_2$. The behavior of $\tau_S$ after hydrogenation cannot currently be explained by the spin relaxation mechanisms proposed for pristine graphene (D'yakonov-Perel or Elliott-Yafet), as in both cases $\tau_S$ decreases with the increase of the spin-orbit coupling.

We would like to acknowledge H. M. de Roosz, J. Holstein and B. H. J. Wolfs for technical support and K. J. Kattukaren for critical reading of the manuscript. This work was financed by NanoNed, the Zernike Institute for Advanced Materials and the Foundation for Fundamental Research on Matter (FOM).

---

* Electronic address: m.wojtaszek@rug.nl

Supplementary information

A.  Spin valve measurements upon hydrogenation. Changes of contact polarization.

To properly resolve the switching of individual Co contacts when sweeping the magnetic field we vary the width of fabricated contacts from 125 nm to 250 nm. The non-local spin valve measurements are always performed using the outermost contacts as a current drain and voltage detector reference. $R_{NL}$ in the fabricated devices shows small (<10%, in case of device B with all ferromagnetic contacts) or non (device A, with non-magnetic outer contacts) extra switches. The presence of only two different resistivity levels in non-local spin valve measurements confirms that the outer detector as well as the current drain do not affect the spin resistivity (no additional switches are observed in spin valve signals).

We observe that upon hydrogenation the contact resistances always increase (by 10-50%) and the polarization of contacts variously changes. Figure 1 shows that the contact polarization can decrease (lower panel) or even switch the polarization (upper panel). Such inversion of contact polarization can be explained by the change of the tunneling interface of the contact during the hydrogenation process, by for example the depletion of oxygen within $AlO_x$ tunneling barrier[1].

In any of the room temperature non-local spin valve measurement (Fig. 1) we do not observe a decrease of $R_{NL}$ around zero magnetic field, which one would expect for uncorrelated magnetic dipole moments originating from hydrogen defects[2,3]. From this we conclude that at the defect concentration estimated here ($4.5 \times 10^{11}$ atoms/cm$^2$) magnetic moments associated with hydrogen defects do not introduce any significant magnetic field or that this effect is randomized by the thermal fluctuations (we note that the spin valve displaying such an effect were measured at low temperatures[2,3], below 15 K).

B.  Fabrication and current spectroscopy of tunneling contacts

The cobalt electrodes are patterned using electron beam lithography and after the development of the resist the layers of Al and Co are evaporated. Before the evaporation of a thin layer of Al for tunneling barrier, when the sample is still screened from the e-beam target by the shutter, we pre-evaporate 10 nm of Ti to lower the base pressure in the chamber ($1 \times 10^{-6}$ mbar). This procedure might deposit a fractional monolayer of Ti on the sample, which would act as a seeding layer for Al and improves the smoothness of the tunneling interface[4]. AFM measurements of the Al film on graphene with and without Ti pre-evaporation, however, show no significant difference in roughness: in both cases the root mean square of surface roughness ranges between 0.4 and 0.6 nm. Therefore we rule out any effect of the Ti pre-evaporation other than the intended reduction in base pressure.

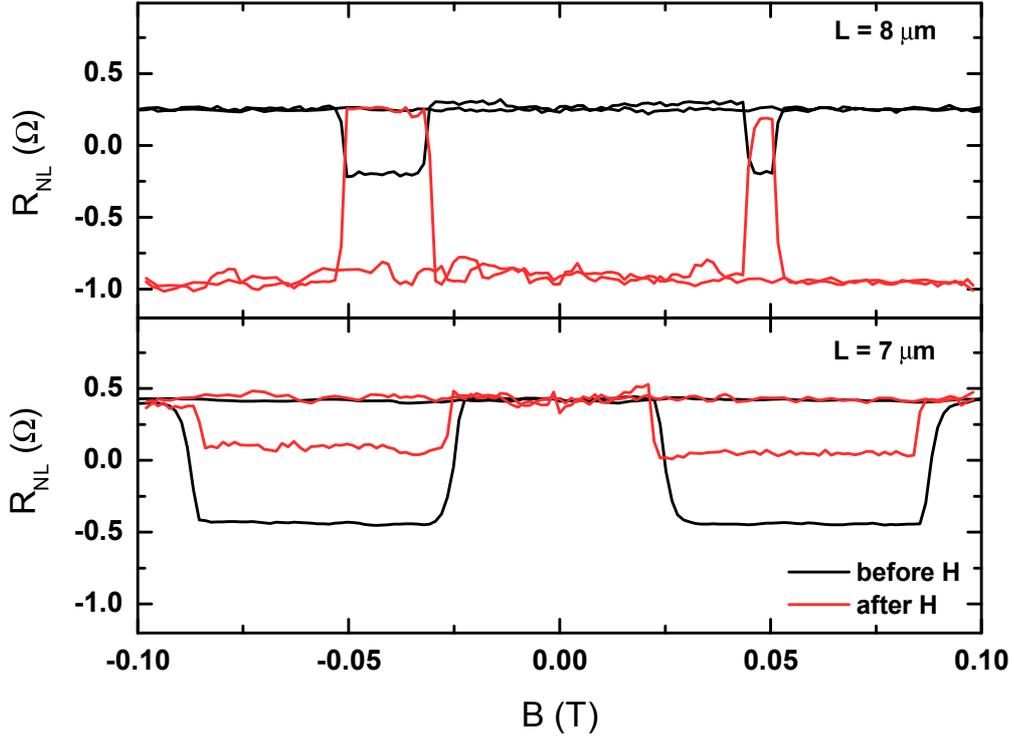

FIG. 1: (Color online) Spin valve measurement in graphene over two different distances before and after hydrogenation (trace and retrace, with the saturation field of 0.5 T) in device B. The hydrogenation process affects contact polarization, decreasing it (bottom panel), increasing it or even changing the sign of polarization (top panel). This can be explained by a change of the interface of the contact, like the depletion of oxygen within the $Al_2O_3$ tunneling barrier during the plasma hydrogenation of the device.

The 3 terminal resistances of the contacts $R_C$ in device A fall in the range from ∼10 to ∼50 kΩ. The optimal range value of contact resistances plays an important role in reducing the conductivity mismatch between a metal electrode and graphene and suppressing the backscattering of spins into the contacts, so that a detectable spin accumulation can be achieved. To quantify that we calculate $R$ parameter $R = R_C W/\rho$, introduced in Ref. 5, where $\rho$ is the graphene resistivity and $W = 1.5$ $\mu$m is the flake width. This parameter represents the strength of the spin relaxation due to the finite contact resistances and indicates the deviations from exponential decay of the spin signal with the distance for $R <$1x10$^{-6}$m. For the contacts in our devices the $R$ parameter ranges from 3x10$^{-6}$m to 1x10$^{-4}$m and explains why we observe non-local signals as high as $R_{NL}$=20 $\Omega$ at a distance of 1.3 $\mu$m.

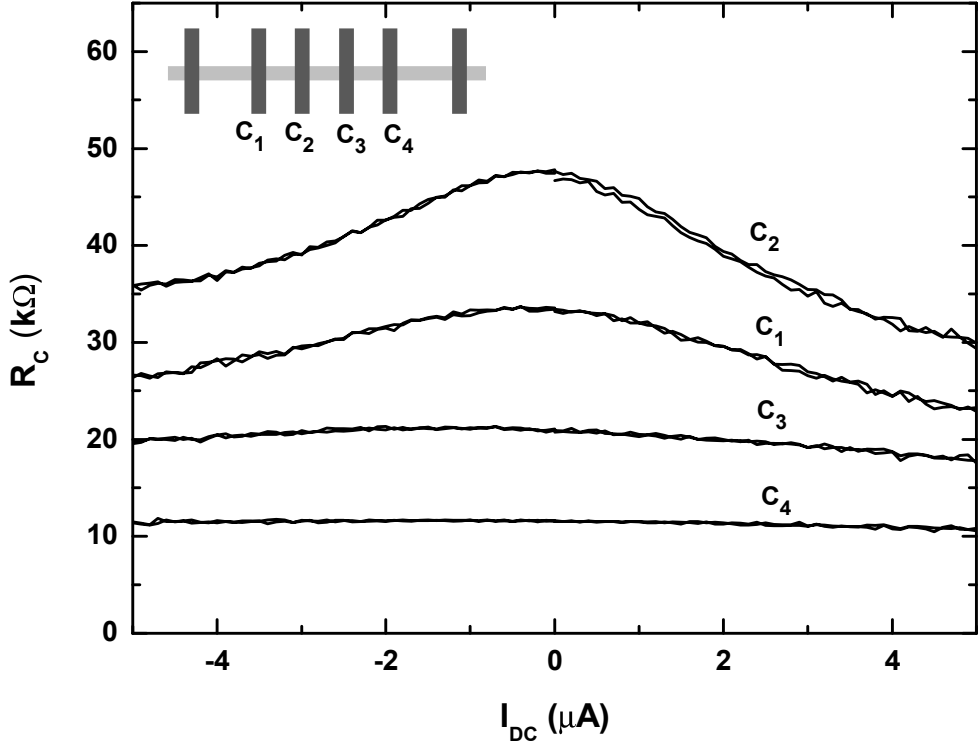

FIG. 2: (Color online) Current spectroscopy of tunneling contacts in a graphene spin-valve device (device A) and their schematic position in the device. All the contact resistances are above 10 k$\Omega$ and their individual values scale with the designed contact area, indicating homogeneous contact resistivity. The contacts display a non-linear change in resistivity with the current bias, most pronounced for the highest resistive contacts. This indicates a tunneling injection through the interface.

To confirm the tunneling character of the contacts, we measure the change of differential contact resistance under DC bias. For that we apply in a 3 terminal configuration, where the source and voltage probe are using the same electrode, a variable DC current $I_{DC}$ with an AC component of 100 nA (to be detected by a Lock-In amplifier). As can be seen in Fig. 2, the non-linear behavior is most significant for the highest resistive contact ($C_2$), though even for the least resistive one ($C_4$) the measurement displays small non-linearity with $I_{DC}$. This indicates the tunneling injection through the ferromagnetic contact and explains high spin signals and spin transport properties ($\tau_S$, $D_S$) in the measured devices.

### C. Comparison between $D_S$ and $D_C$ with broadened density of states.

The diffusion coefficient can be independently extracted from charge ($D_C$) and from spin ($D_S$) transport. $D_C$ is calculated using the Einstein relation $\sigma = e\,\nu(E)D_C$, where $\nu(E)$ is the density of states and $\sigma = 1/\rho$ is the graphene sheet conductivity. In the theoretical limit of T = 0 K the number of states at the Dirac point in ideal graphnene vanishes and that leads to a singularity of $D_C$ at $V_D$. This can be eliminated by introducing a Gaussian broadening to $\nu(E)$, which collectively accounts for finite temperature, electron-hole puddles and possibly finite lifetime of electronic states[6]. In our case, the good agreement between $D_C$ and $D_S$ before hydrogenation is obtained when $\tilde{\sigma} = 110$ meV, see Fig. 3, and is close to the values reported in the literature[3,6]. Though with the same value of $\tilde{\sigma}$ one can qualitatively imitate the behaviour of $D_S$ also after hydrogenation, we observe a slight offset between $D_C$ and $D_S$ of 0.002 m$^2$/s,, which cannot be compensated by a further increase of $\tilde{\sigma}$. In the sparse low temperature data (T = 70 K for device A, T = 4K for device B) we also do not observe any significant discrepancy between the values of $D_C$ and $D_S$.

Such a discrepancy can originate from the presence of localized states in graphene or its close vicinity[7] or by the modulation of the gyromagnetic ratio due to the effective magnetic field from magnetic moments induced by hydrogen defects[3]. The latter case does not apply here because at room temperature the thermal fluctuations randomize these localized magnetic moments. As we do not observe a dip in spin-valve measurements of $R_{NL}$ at low magnetic field neither the difference between $D_C$ and $D_S$ we do not expect a change in the $g$-factor. Following the discussion in Ref. 3 the eventual enhancement of the $g$-factor at room temperature is estimated to be less than 2 %, therefore it is reasonable to use the free-electron value $g = 2$ in Hanle fits. From that we conclude that the probable reason for a small offset of $D_S$ with respect to $D_C$ after hydrogenation is the modulation of the density of states due to localized states. Nevertheless, the general similarity between $D_S$ (from Hanle fits) and $D_C$ (calculated using the standard graphene DOS) confirms that the DOS in softly

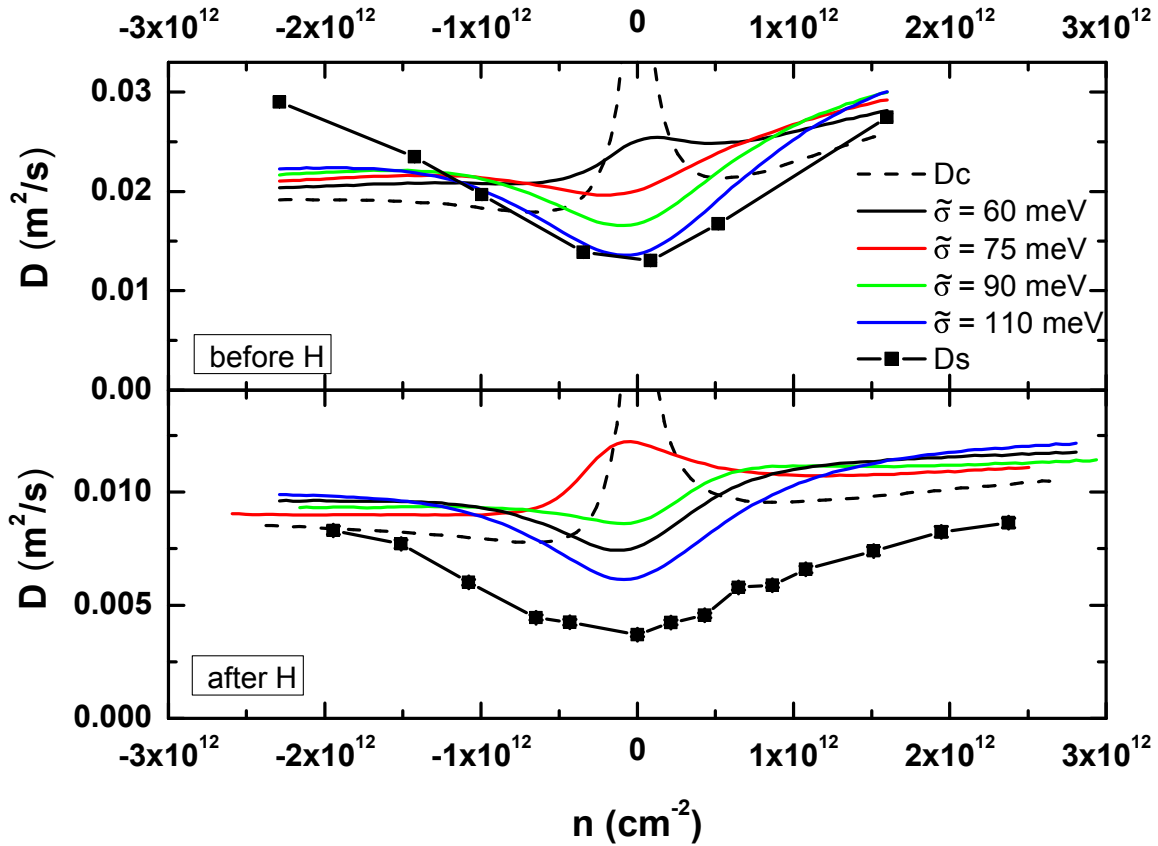

FIG. 3: (Color online) Comparison of the diffusion coefficient extracted from charge-$D_C$ and from spin-$D_S$ transport measurements, before and after hydrogenation for device A. Dashed lines display $D_C$ calculated for density of states $\nu(E)$ without any broadening, continues lines display $D_C$ obtained when a Gaussian broadening of $\tilde{\sigma}$ is introduced to the $\nu(E)$. The square points represent the diffusion coefficient obtained from Hanle precession measurements ($D_S$). A good agreement between $D_C$ and $D_S$ is obtained for $\tilde{\sigma} = 110$ meV for the case before hydrogenation. After hydrogenation there is an offset between $D_C$ and $D_S$ of 0.002 m$^2$/s, which cannot be compensated by a further increase of $\tilde{\sigma}$.

hydrogenated graphene is not significantly affected.

### D. Hanle measurements for different distances between injector and detector.

We present the details of the elementary fits and extracted coefficients for different distances $L$ between injector and detector. In device A the measured spacings are $L$=1.3, 3.3 and 5.5 µm; in device B $L$= 1.5, 3, 4, 6.5, 7, 8 and 11 µm. In Fig. 4 and Fig. 5 we present three representative regions of each device. The obtained spin coefficients are very similar within the individual set (before or after hydrogenation). In most of the data we apply the fitting procedure to the difference between the Hanle precession measurements at parallel and antiparallel magnetizations to keep the fitting offset at zero.

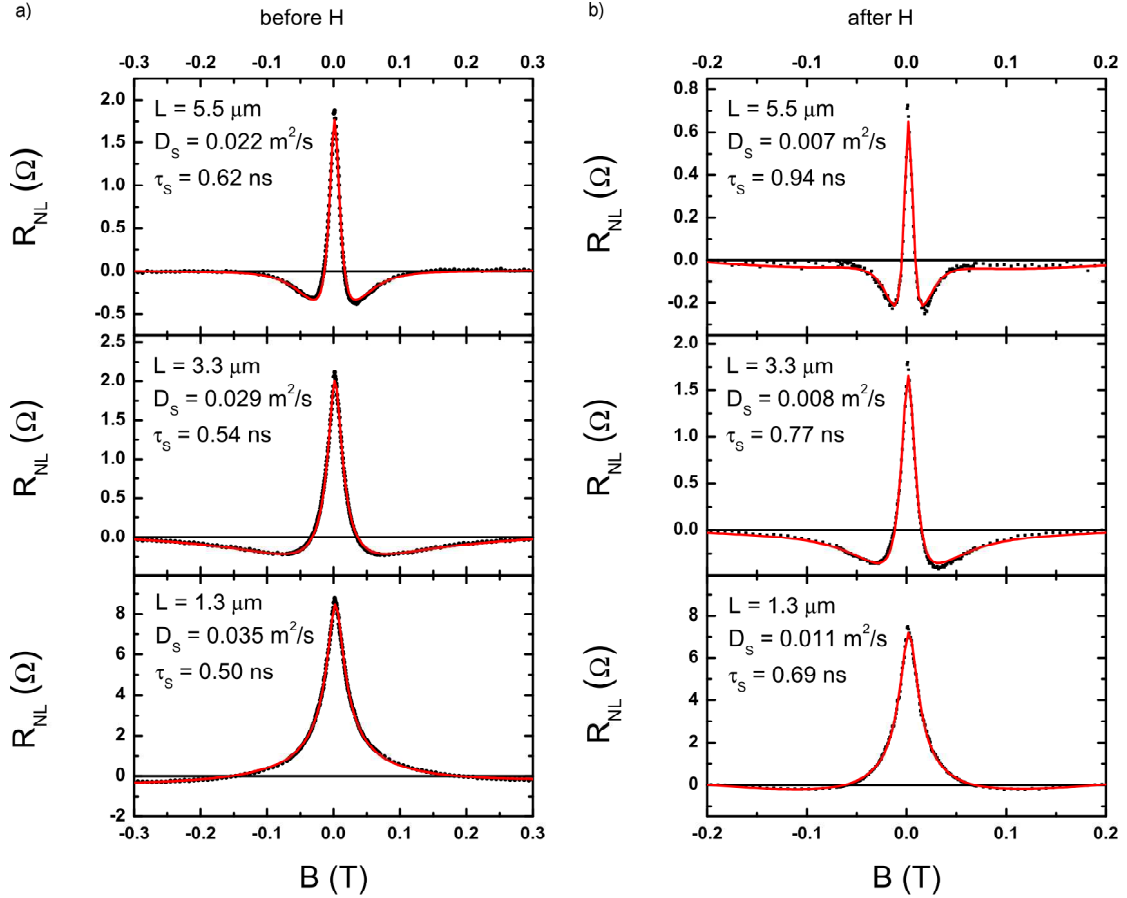

FIG. 4: (Color online) Hanle precession curves of device A for three different distances $L$ between injector and detector, (a) before hydrogenation, (b) after hydrogenation. Red line represents the fits of the data to Bloch equation and extracted spin coefficients are displayed next to the corresponding data. All presented data are obtained for carrier concentration $n = 2.2 \times 10^{12}$ cm$^{-2}$ ($V_g - V_D$ = 50 V). Notice the different scales of the magnetic field before and after hydrogenation.

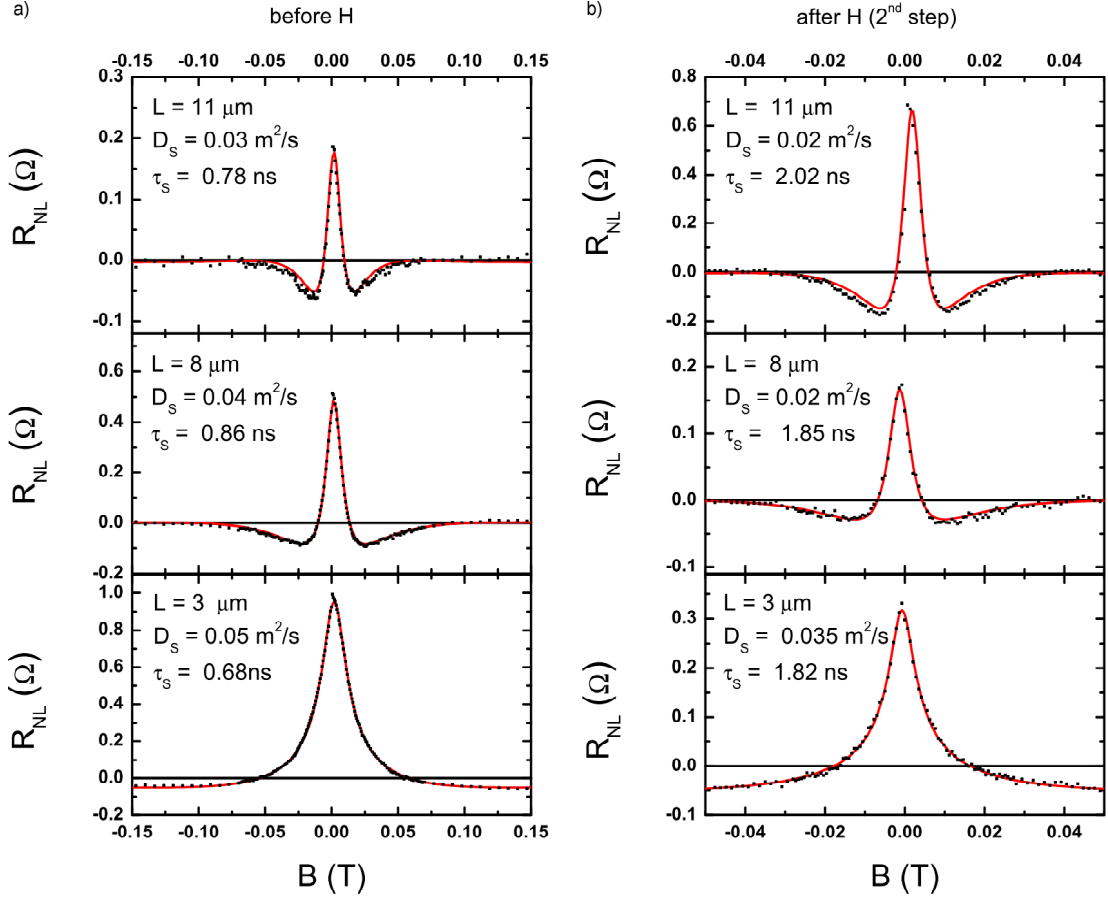

FIG. 5: (Color online) Hanle precession curves of device B for three different distances $L$ between injector and detector, (a) before hydrogenation, (b) after $2^{nd}$ hydrogenation step. Red line represents the fits of the data to Bloch equation and extracted spin coefficients are displayed next to the corresponding data. All presented data are obtained for carrier concentration $n = 2.2 \times 10^{12}$ cm$^{-2}$. Notice the different scales of the magnetic field before and after hydrogenation.